\documentclass[a4paper,11pt,article,oneside]{memoir}
\usepackage{ecis2019}

\usepackage{graphicx} 
\usepackage{csquotes}
\usepackage{enumitem}
\usepackage[center]{caption}

\maintitle{Tell Me How You Feel: Designing Emotion-Aware Voicebots to Ease Pandemic Anxiety In Aging Citizens}

\shorttitle{Tell Me How You Feel}

\authors{
Mieleszczenko-Kowszewicz, Wiktoria, SWPS University of Social Sciences and Humanities, Warsaw, Poland, s20024@pjwstk.edu.pl

Warpechowski, Kamil, Polish-Japanese Academy of Information Technology, Warsaw, Poland, k.warpechowski@pja.edu.pl

Zieliński, Kazimierz, Polish-Japanese Academy of Information Technology, Warsaw, Poland, k.zielinski@pja.edu.pl

Nielek, Radosław, Polish-Japanese Academy of Information Technology, Warsaw, Poland, nielek@pja.edu.pl

Wierzbicki, Adam, Polish-Japanese Academy of Information Technology, Warsaw, Poland, adamw@pja.edu.pl
}

\shortauthors{Mieleszczenko-Kowszewicz, et al.} % ← This goes into the header. 

\begin{document}
%%
%% The abstract is a short summary of the work to be presented in the
%% article.
\vspace{-1.5cm}
\begin{center}
%\textit{Research Paper}
\end{center}
\vspace{1cm}
\begin{abstract}\noindent
The feeling of anxiety and loneliness among aging population has been recently amplified by the COVID-19 related lockdowns. Emotion-aware multimodal bot application combining voice and visual interface was developed to address the problem in the group of older citizens. The application is novel as it combines three main modules: information, emotion selection and psychological intervention, with the aim of improving human well-being. The preliminary study with target group confirmed that multimodality improves usability and that the information module is essential for participating in a psychological intervention. The solution is universal and can also be applied to areas not directly related to COVID-19 pandemic.
\end{abstract}


\chapter{Introduction}

In early 2020 mankind was challenged by a new natural disaster: COVID-19. The pandemic resulted in several innovations, such as the new vaccines, but those also in the field of information technology. Technology creators and users realized the need for adaptation and further development of several Web-based and mobile technologies used for distance work, education, or healthcare. One of the early responses to the spread of the virus was the development or extension of chatbots that have functions helping people during the pandemic. Among main goals of these efforts was providing reliable information about the disease and supporting initial self-diagnosis.

This article refers to one of severe side-effects of the pandemic: the influence of enforced isolation on human well-being and emotional health. In particular, we have been interested in reducing feelings of anxiety and loneliness caused by isolation. Research shows (\cite{choi2020depression, gonzalez2020mental}) that one of consequences of COVID-19 was experiencing coronavirus-related anxiety. Isolated persons also experienced hopelessness, suicidal ideation, spiritual crisis and used maladaptive coping strategies, such as use of drugs or alcohol (\cite{lee2020coronavirus}). Many countries, like China (\cite{liu2020online}) or Indonesia (\cite{ifdil2020online}), introduced technological solutions to improve citizens' emotional health.

In order to achieve this goal, we planned to create a new technology that would be as accessible and as easy to use as possible. We considered any interface limitations or barriers as an obstacle to our goal, since they could result in an increase of user frustration instead of relieving anxiety. We needed a technological solution that would not only allow the chatbot to communicate using text or voice, but also use the well-known and intuitive touchscreen interface. Chatbots using our technology can run on inexpensive (legacy) tablets that are widely available today.

The basis of our solution is, therefore, a multimodal chatbot technology supporting several unique functions. The first step towards improving a person's emotional health is to make the person aware of their own emotions. For this reason, our chatbot - Rita - allows the user to express their emotions. Rita can use this knowledge in the next stage of her interaction with the user in ACT therapy. To address the need of users for reliable information about the pandemic, we implemented in Rita an informational function. 

We have tested Rita in a qualitative experiment with users isolated during lockdown. Our results are both informative and surprising. The informational function of Rita turned out to be a key element that enabled users' trust and supported their engagement in therapy. The voice interface also turned out to be a key feature that helped users to treat Rita as (almost) human, and, therefore, improved users' reaction to therapy. 

The rest of this article is structured as follows. The next section discusses related work. Section 3 describes Rita's design. Section 4 introduces experiment design and measures, while Section 5 discusses experiment results. Section 6 concludes the article.

\chapter{Related work}
Chatbots provide space for a new and more natural form of communication with a computer (\cite{brandtzaeg2017people}).  \citet{jain2018evaluating} and \citet{liao2018all} proposed that bot interactions should exhibit social skills. Bots should communicate with accessible, friendly, and natural language (\cite{folstad2017chatbots}). Chatbots should not be as humanoid as possible because the effect of the valley of weirdness causes fear, anxiety, and increased expectations (\cite{ciechanowski2019shades}).
The most critical aspect of chatbot design is the information about errors and unexpected user input (\cite{fichter2017chatbots}). %\citet{gnewuch2018faster} proposed that simulation of delays make conversation more natural, because are similar that human-like interaction. 
The visual aspect is becoming more and more significant. Typeface or graphic interface may motivate user to frequent interaction (\cite{candello2017typefaces}).
Voice conversational agents showed great potential for regular users, people with disabilities and visual impairments (\cite{pradhan2018accessibility}) and older adults (\cite{kowalski2019older}). Virtual assistants often increase well-being, health monitoring (\cite{maharjan2019hear}) and are helpful for older adults (\cite{sayago2019voice}).

%\vspace{-.2cm}
\section{Multimodal bots}

Multimodal chatbots receive input and respond by various communication channels, e.g., text, voice, graphic, facial expression (\cite{knote2018and}). Multimodal interfaces are becoming more engaging, also for people with disabilities (\cite{brunete2021smart}). Joining VUI, GUI, and other interaction methods make a product more inclusive, valuable, and attractive for end-users. The most popularized sectors are Smart TV (\cite{fernandes2018review}), automobiles (\cite{meng2020voice}) and education (\cite{ruan2019bookbuddy})

Attractive multimodal interfaces are intelligent speakers with touch screens (Google Nest Hub, Amazon Echo Show) and smartphones or smartwatches with a built-in assistant (Siri, Google Assistant). They create visual responses complementary to voice communication. However, interactions are an integral part of operation systems. Developers can use only a few predefined response types (\cite{skorupska2020conversational}). It is often not insufficient for bot designers.

Multimodal interfaces may also use gestures and GUI (\cite{brunete2021smart}). \citet{kepuska2018next} proposed the hybrid interface that combines voice, video, gestures, graphics, gaze, and body movement.

%\vspace{-.2cm}
\section{Emotions in chatbot conversations}
People feel more comfortable communicating with a chatbot when it exhibits emotions (\cite{ho2018psychological}).
Personalized chatbot communication (in example:  speed speech,  use of user names, generate responses based on user well-being) increase perception of the human-computer interaction (\cite{andrews2012system}). Chatbots should have social skills. It is essential for natural conversations (\cite{ciechanowski2017necessity}).
The bot will be empathetic when it can detect user emotions (\cite{oh2017chatbot}). Software may measure emotional state implicitly by analyzing text input and audio  (\cite{zhou2020design}) or directly by visual psychological questionnaires (\cite{plutchik1980general}).

\section{Psychological therapy and intervention}
Psychological interventions using chatbots are not new. The first computer application simulating conversation with psychotherapist was Eliza (\cite{weizenbaum1966eliza}). The idea was simple - software rephrasing sentence wrote by user and replying it back as a question. Conversation with Eliza resembled those with a psychologist, and it was based on Rogerian client-center therapy. 

Advances in natural language processing have contributed to the emergence of an increasing number of chatbots applying different therapy's approach. 
The Woebot chatbot is automated conversational agent based on cognitive behavioral therapy. It has a therapeutic effect consisting in engaging user with short conversation and measuring emotion. Research shows that using Woebot by depressed patients results in a significant reduction of symptoms. (\cite{fitzpatrick2017delivering}).

\citet{ho2018psychological} show that making an emotional disclosure to a chatbot brings similar positive effect as when disclosing emotions to people. 
The areas of potential use of chatbots in psychological interventions are  broad: from providing clues for well-being and self help guidance (\cite{cameron2018best}), to more specialized chatbots enhancing the effectiveness of communication with people with autism disorder (\cite{mujeeb2017aquabot}). Emergence of COVID-19 created a new areas of chatbot's application: easing anxiety caused by different concerns connected with this disease.

%\vspace{-.2cm}
\section{Bots designed to cope with pandemic}
Since the outbreak of the pandemic a number of chatbots have been designed with the purpose of coping with various aspects of it, and some existing popular chatbots were amended with new functionalities.
COVID-19 chatbots were applied to fulfill the following goals:

\begin{itemize}
    \item disseminating information about symptoms, medication, precautionary measures, advice on best practices for prevention
    \item combating misinformation and fake news about COVID-19 by providing proven factual information
    \item self-checking (self-triage), symptom monitoring, personal risk assessment, behavior change support 
    \item monitoring social contacts, exposure to COVID-19, providing notifications to individuals or authorities, following up high-risk groups
    \item tracking physical and mental health during quarantine and isolation, mitigating psychological symptoms (anxiety, distress, depression)
\end{itemize}

The US Centers for Disease Control and Prevention (CDC) created a self-checker bot named Clara\footnote{\url{https://www.cdc.gov/coronavirus/2019-ncov/symptoms-testing/coronavirus-self-checker.html}} and another one named "COVID-19 Viral Testing Tool"\footnote{\url{https://www.cdc.gov/coronavirus/2019-ncov/testing/index.html}} to aid decision-making after COVID-19 viral testing.
Apple updated its voice assistant Siri to include a self-checking questionnaire\footnote{\url{https://www.cnbc.com/2020/03/21/apple-updated-siri-to-help-people-who-ask-if-they-have-coronavirus.html}}. 
WHO launched a Facebook Messenger based chatbot with the main purpose to combat COVID-19 misinformation\footnote{\url{https://www.who.int/news-room/feature-stories/detail/who-launches-a-chatbot-powered-facebook-} \url{messenger-to-combat-covid-19-misinformation}}, which it did by specifically addressing common myths and providing the appropriate clarification to any such a myth. Li et al. designed and built Jennifer – a chatbot aiming at combating misinformation, maintained by a global group of volunteers (\cite{li2020jennifer}).

After these initial attempts many governments around the globe followed with chatbot-based national solutions for general coronavirus information spreading, self-checking of symptoms, for testing and vaccination\footnote{\url{https://www.messengerpeople.com/governments-worldwide-covid-19/}}. 

UCSF Health designed a chatbot-based workflow to screen healthcare workers for COVID-19 symptoms and exposures prior to every clinical shift (\cite{judson2020implementation}).
\citet{battineni2020ai} proposed an artificial intelligence chatbot for the purpose of diagnostic evaluation and recommending immediate measures when patients were exposed to COVID-19.
Some screening chatbots were designed with the specific aim of reducing the strain on the health care system, by detecting possible cases of COVID-19 from the patient's conversation with the bot (\cite{erazo2020chatbot}).

A holistic approach of personalized digital coaching in a variety of life topics impacted by COVID-19 related restrictions was taken by ELENA (\cite{ollier2021face}). The chatbot keeps the user up to date with COVID-19 related guidelines while providing mental health support and assistance for physical activities. \par
To the best of our knowledge, an implementation of the ACT therapy with the support of a comprehensive information module, using a multimodal chatbot technology, with low profile hardware requirements on the user side, has not yet been described.

\begin{figure}[t]
\begin{center}
  \includegraphics[width=8.1cm]{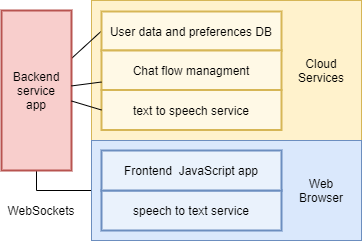}
  \caption{Rita chatbot infrastructure}
  \label{fig:infrastructure}
    %\vspace{-.3cm}
\end{center}
\end{figure}

\chapter{Rita's design}
Our chatbot has a modular design. Each module in Figure \ref{fig:my_label} is agnostic of the design of other modules. Potential developers can replace any module with another technology. Modules communicate by standard HTTP.
The architecture of information was split into blocks. Dialogflow application or similar system allows for developing conversation with many people at the same time. Rita recognizes a few general sentences in each place in the system, but an author can define specialized sentences in each module. Users can select any module directly using voice or a visual menu.

%\vspace{-.2cm}
\section{Informing about COVID-19}
Rita's information module provides verified news about the COVID-19 pandemic. The information is divided into sections following the structure proposed by WHO. 
Information module is divided into the following sections:
\begin{itemize}
    \item Facts and myths about COVID-19 
    \item Rehabilitation after COVID-19
    \item Stress in isolation
    \item Tutorial on usage of masks and disinfectants
\end{itemize}
The section devoted to correcting common myths has an additional feedback option, where the user is asked whether the last information (i.e. a correction of a myth) was regarded by him as helpful or not.

\begin{figure}[t]
    \centering
    \includegraphics[width=7.5cm]{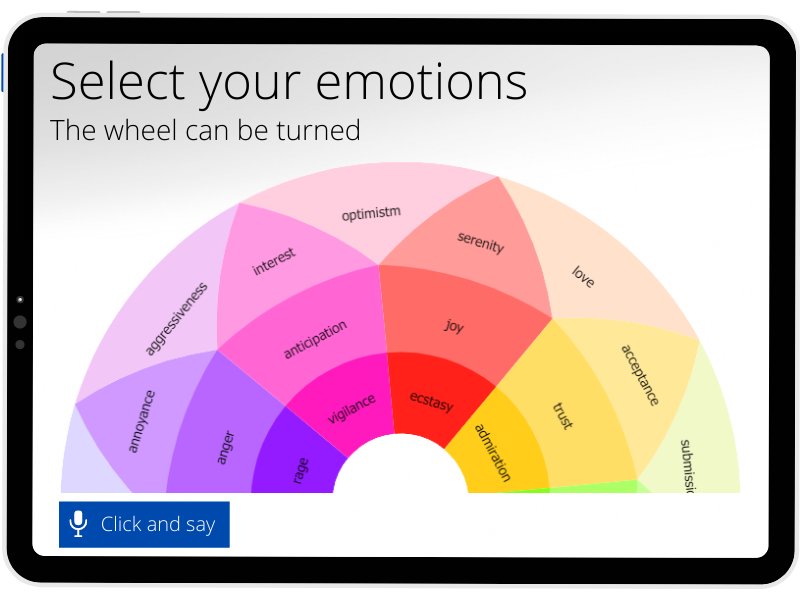}
    \caption{Emotion Wheel on Rita}
    \label{fig:screen}
    %\vspace{-.6cm}
\end{figure}

%%\vspace{-.2cm}
\section{Measuring emotions}
Figure \ref{fig:screen} presents the Plutchik Emotion Wheel (\cite{plutchik1980general}) optimized for touch interfaces (\cite{warpechowski2019tagging}). Users may select emotions on screen or use voice. The system recognizes user speech and automatically selects emotions on GUI. Feedback is vital in voice interfaces. Users should be informed about bot actions or mistakes.
This type of emotion self-assessment allows for storing structured and comparable data. Emotions are difficult to describe for many people. For them, the display is helpful in learning names that can denote their emotions. Experienced users can select emotions by voice - often, this is the fastest way. GUI and VUI are complementary.

%%\vspace{-.2cm}
\section{Therapy intervention}
Therapy intervention is Rita's third module that aims at decreasing the anxiety caused by coronavirus outbreak among people who not only suffered from this disease but also were afraid of infection.

This module comprises 5 steps based on Hariss (\cite{harris2020face}) set of practical steps of responding to the Corona crisis and uses principles of acceptance and commitment therapy (ACT).

The main idea of this therapy is that humans make subjective associations, called “cognitive fusion”, between different stimuli. Individuals are more strongly influenced by those associations represented by linguistic representations of events, than by their own present experience (which they cannot focus on). Negative emotions connected with linguistic representations can cause mental pain. The role of the therapist in ACT therapy is to make the patient accept those feelings and to solve psychological problems by increasing the key skill of psychological flexibility. As a result patients can fully experience the present moment and adjust their behavior, so that it fits their own values.  (\cite{wang2017acceptance})
 
In Rita, the first therapeutic insight starts in the information module where users can declare which values are important to them (Fig.3). Rita asks users to choose from a checklist the values that are important to them: family, work, health etc. This question not only builds user's profile in the application, but also encourages the user to think about what is important for them, and, as a consequences, develops self-consciousness. 

User enters the therapy module after declaring their emotions in the measuring emotions module. The therapy intervention is a finite state-based system in which the user is guided through the module step by step to achieve a goal (\cite{mctear2002spoken}). Each step in the therapy module involves a different processes consistent with the ACT therapy. 

The key core processes that are crucial in ACT therapy are: (1) acceptance, consisting in presenting a positive attitude of embracing any experience; (2) cognitive defusion, consisting in distancing and separating oneself from one's thoughts and feelings in such a way as to see them as they really are (words constructing narratives and passing impressions), and not as they appear to be (threats or facts); (3) being present, consisting in consciously experiencing the present moment without evaluating it;  (4) self as context, consisting in shifting patient's perspective from "self as perspective" to " self as a context";  (5) values, which are qualities that define directions of actions. Following a value helps to live meaningful lives. (6) Committed action, consisting in helping patients to undertake actions consistent with their values.  

Rita's therapy module is a dialogue initiative module in which the system leads the conversation (\cite{laranjo2018conversational}). Each step provides information about psychological mechanisms and instructions on what one should do to reduce anxiety connected with coronavirus' outbreak. Moving to the next stage of therapy requires an action from the user, such as verbal confirmation of understanding what the bot has said.

The therapy module uses data provided by the user in previous modules, such as gender, values, declared emotions on emotion's wheel and the user's verbalised threatening thoughts connected with the coronavirus. Gender is used to personalize the communication with proper inflectional endings. Declared emotion and thoughts that caused the emotion are used to narrow the field of work in therapy. Finally, the declared value serves to propose an activity that can reduce anxiety. 

\begin{figure*}
    \centering
    \includegraphics[width=\textwidth]{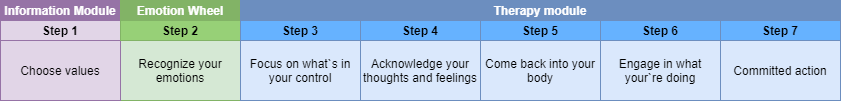}
    \caption{Therapy steps in Rita's modules}
    \label{fig:my_label}
\end{figure*}

%\vspace{-.2cm}
\section{Touch interface for voice bot}
\paragraph{\normalfont Graphical User Interface for voice expands functional possibilities.}
During the first interaction using VUI, a user should learn allowed commands and bot options. Authors of virtual assistants create documentation with command specifications. GUI reduces the cost of onboarding as it directly shows interaction suggestions \footnote{\url{https://www.cnet.com/home/smart-home/every-alexa-command-you-can-give-your-amazon-echo-smart-} \url{speaker-or-display/}}.
When used frequently alone, VUI can be tiring. The bot must say all information each time. However, in the hybrid interface, voice can be used for content summarization, while GUI displays complete data.
The screen reduces the navigation time for advanced users. User does not have to wait to listen to the entire message. They can stop reading using the graphical interface button to move on to the next step of a dialogue.
The touch method of human-computer interaction should be complementary to VUI. All steps and functionalities should be feasible independently of the communication method. Rita's universal design extends the scope of Rita's usage. Users may continue the dialogue in a public area with privacy. Multimodal communication is more inclusive than VUI. Users may select the method of communication suitable for each situation and their own abilities. People with vision problems may use VUI more often. 

Multimodal assistant allows for continuous usage design. On GUI, display notifications are possible. User should be focused only on the direct usage (for example, a reminder may not be heard, a user may be remote from the assistant). Screens and other visual communication allow for generating asynchronous notifications.

Voice Assistants available on the market allow for initializing communication only from end-user. The developer cannot build an application that starts dialogue from the bot at any time. The proposed multimodal interaction resolves this problem using methods that engage various modalities.

Each of the GUI responses requires models developed in code. To reduce development time, we have created a few helpful templates. Modifications and new patterns require the release of new versions, but the designer controls text or other placeholders without any code changes. Each of the templates has implemented default voice interactions and graphic CTA.
\begin{itemize}
  \item Slides - boxes without sample commands
  \item Checkboxes - multiple choice list
  \item Emotions - emotion selector based on wheel (\cite{warpechowski2019tagging})
  \item Dashboard - weather, COVID-19 information, CTA
\end{itemize}

The rest of the responses is based on the default template. GUI interaction may contain header, text, or HTML frame (for example, images, web content) and buttons with command suggestions. Designer may decide which text is additionally said by the assistant's voice. By default, all information presented on GUI is the same as VUI.

Rita is developed for multiple usages by the same user. Rita interprets responses and stores personal information for each user. Chatbot on the start session generates personalized greetings with a user name. In any communication, the designer can use variables based on stored data. In the experiment, it was necessary to recognize user's gender for word inflection. A part of the psychological module is based on previous user responses. 
Users feel better and talk longer when the bot generates personalized interactions (\cite{shumanov2021making}).

%\vspace{-.2cm}
\section{Technology stack and hardware}

Rita uses any web browser as a core environment. Progressive Web Apps  allow for creating a fullscreen application with built-in voice-to-text recognition. Scenarios and dialogue options are defined in Dialogflow CX. It is a new implementation of Google Dialogflow dedicated to creating virtual agents. It has an easy user interface for creating data block dialogues without any code. It allows for working only with text, but it accepts to create customization JSON responses. Rita has the backend service written in NodeJS localized in the exact server location in Google Cloud, which converts JSON to rich responses. Backend simultaneously uses Google Cloud for text transformation to audio. In JSON schema it is possible to use SSML language to create specific voice responses. The backend uses user data stored in the persistent database to create personalized communications. User interface required two-way communication with reduced delays.

\chapter{Experiment Design}

%\vspace{-.2cm}
\section{Research questions}
Rita is a hybrid bot, as it can listen to voice or GUI inputs, and it uses texts, graphics and voice for outputs. As such, we expect it to be more involving for the user than a bot that uses only voice, only text, or only GUI. Hence our first question:
\begin{enumerate}
    \item Is the hybrid user interface more engaging than a voice interface or visual interface alone?
\end{enumerate}
In addition, we wanted to examine users' perceptions on further applications of Rita's hybrid interface:
\begin{enumerate}[resume]
    \item Can Rita become a daily assistant?
    \item Should Rita display information on a continuous basis?
    \item Should Rita initiate conversations?
\end{enumerate}
Furthermore, we wanted to see whether verified information provided by Rita would be perceived as credible: 
\begin{enumerate}[resume]
    \item Which sources of information are regarded as credible by the users?
    \item Are users aware of misinformation regarding the pandemic?
    \item Will users perceive information provided by Rita as credible and reliable?
\end{enumerate}
Finally, we wanted to see users' feedback on the ACT therapy implemented in Rita: 
\begin{enumerate}[resume]
    \item Can Rita improve a person's well-being during the pandemic?
    \item What makes a person engage in Rita's therapeutic tasks? 
    \item Which features of chatbots/voicebots will make people use them more willingly as a tool to improve one's well-being?
\end{enumerate}

%\vspace{-.2cm}
\section{Participants}
The criteria of inclusion was being isolated from people during epidemic. Participants were both people who were working and retired citizens (N=10). The subjects' anxiety score average was 3.0 (SD=0.66) and helplessness score average was 2.47 (SD=0.62). We also assumed that another criterion should be age, due to the fact that people above 40 years old were more threatened with coronavirus' consequences (\cite{yang2021estimating}). Participants were recruited among friends and family of the research team.  A total number of participant was 10 (5 female, 5 male).

The mean age of the subjects was 59.3 years (SD=12.0). 

According to the \emph{internet efficacy level} (described in \ref{subsection:efficacy}), one participant was a "light" Internet user, 4 participants were "medium" Internet users, and 5 participants were "heavy" Internet users. 

%\vspace{-.2cm}
\section{Procedure}

Participants took part in the procedure which comprised with two stages: in-depth interview and task-oriented part. \\
During interview participant answered questions about (1) credibility of information during COVID-19, (2) coronavirus impact on life and mental health. 

In task-oriented part users performed five different tasks: (1) Personalization of multimodal bot (2) Verification of coronavirus fake news, (3) Finding the advice for people who have a contact with someone with COVID-19, (4) Finding information about rehabilitation. (5) Participation in the therapy inside application. 

Each test session took approximately 28 minutes. Participants were verbally instructed which task they should do. Researcher did not suggest what type of modality participant could choose (spoken, visual, mixed). Interviews were audio-recorded and transcribed and the logs from the task were downloaded. 

Both participants and researchers wore medical masks to avoid potential coronavirus infection.
Interviews were recorded and transcribed. Researcher reread the following material and coded it by piece of text related to answer for research's questions. 

The procedure was reviewed and approved by the Ethical Commission on Polish Japanese Institute of Information Technology. All subject gave written informed consent in accordance with the Declaration of Helsinki. Both subjects and experimenter wore masks during interview to avoid potential risk of COVID-19 contagion. 

%\vspace{-.2cm}
\section{Measures}

\textbf{Session Evaluation Questionnaires}\\
Respondents rated a single therapy session on four dimensions: 
\begin{itemize}
    \item Depth - which refers to the perceived power and value of the session
    \item Fluency - which refers to the comfort, relaxation, and enjoyment of participating in the session
    \item Positivity - defined as a sense of confidence and clarity, as well as happiness and happiness, and absence of fear or anger.
    \item Arousal - refers to a sense of activity and excitement as opposed to quiet and calm.
\end{itemize}
The first two dimensions relate to the subjects' perception of the session and the last two dimensions relate to the impact and value of the session to the subject. 
The mean scores of depth was M=5.14, (SD=1.187), fluency M=5.95, (SD=0.78), positivity M=6.11 (SD=0.83), and arousal M=3.67 (SD=0.95). The first three scores are relatively high, but the last one is low, which may result in insufficient therapeutic impact.

%\vspace{-.2cm}
\section{Usability Metrics}
System Usability Scale (\cite{brooke1996sus}) is the most popular and fast questionnaire for usability testing. It contains ten rating items with five options of response. The maximum result is 100 points. However, systems with 68 points or more are classified as above-average usability systems.
User Experience Questionnaire (\cite{laugwitz2008construction}) allows for collecting ratings with six dimensions: Attractiveness, Perspicuity, Efficiency, Dependability, Stimulation, Novelty. 
Authors prepared the application for a benchmark with other systems. It includes statistics from around 20 thousand persons based on 452 software products.

%\vspace{-.2cm}
\section{Internet efficacy}
\label{subsection:efficacy}
We measured Internet efficacy of the participants using a metric combining the Web-Use Skill Index (\cite{hargittai2012succinct}) with a number of activities on the Internet, which are performed at least once a month (\cite{kkakol2013subjectivity}). Hence, this metric takes into account both the user's knowledge about the Web and their experience in using it. The measure groups the participants into 3 groups: light, medium and heavy users. 

\chapter{Results}

\begin{table}
\label{tab:ueq}
\centering
  \begin{tabular}{lcl}
    \toprule
    Scale&Mean&Interpretation\\
    \midrule
    Attractiveness & 1,91 & In the range of the 10\% best results  \\
    Perspicuity    & 2,16 & In the range of the 10\% best results \\
    Efficiency     & 1,70 & 10\% of results better, 75\% of results worse \\
    Dependability  & 1,77 & In the range of the 10\% best results         \\
Stimulation    & 1,39 & 10\% of results better, 75\% of results worse \\
Novelty        & 0,89 & 25\% of results better, 50\% of results worse \\
  \bottomrule
  \end{tabular}
%\captionsetup[table]{skip=0pt,singlelinecheck=off}
\caption{User Experience Questionnaire results}
\end{table}
Participants focused on system perspicuity. It is the best-rated property in UEQ results (Table 1). The multimodal interface is attractive and effective. The average value of the System Usability Score is 79/100.
Everyone started interactions via voice. They usually switched to touch if the system did not correctly recognizevoice commands or they had no idea about the following command. Users often used buttons for system navigation (back to the main screen). No one used the optional button for repeat interaction. 

The respondents appreciated the cooperation between voice and screen. They felt confident that the Bot understood their commands because they saw request and responses on the screen.
One of the users was reading faster than listen. He preferred a short interaction without waiting for the end of Bot's voice. He suggested that voice speed should be personalized based and previous interactions.

Respondents said that the application was engaging. The primary values were not getting lost and predictability of interactions.

%\vspace{-.2cm}
\section{Sources of information related to COVID-19}
Participants were interested in the development of the pandemic, and, hence, they followed COVID-19 related news with interest. Every user had their favourite sources of information. Responses ranged from general TV broadcast news to specific information services available on internet. Only one person mentioned radio broadcasts and a printed newspaper as an information source. 

Polish government runs a patient-oriented service dedicated to the matters of public health (\url{pacjent.gov.pl}), which features a section devoted to COVID-19. That section delivers COVID-19 related news and also addresses some common myths. Our research participants were generally aware of existence of such a service, however, nobody had used it on regular basis, with exception of one participant, for whom \url{pacjent.gov.pl} was the primary source of information about the pandemic.

It was assumed that favourite sources of information of all users were generally regarded as credible by each user.

%\vspace{-.2cm}
\section{Misinformation about COVID-19}
Two participants were not aware of the existence of any false news about the pandemic. Remaining participants perceived some information as intentionally false:

\begin{displayquote}
I heard the news of a corporation with a global activity, whose aim is to reduce the population world wide.... To me it is of low credibility. A little scary, but I don't want to believe it. (M66)
\end{displayquote}
% WMK1
This utterance shows that in order to assess a particular false news, the participant was relying both on his instinctive feeling and his internal will not to become scared. 

\begin{displayquote}It is a hyped up flu (laughing). For political or commercial purposes. One can often hear that the pharma industry created this virus on purpose. (M86) \end{displayquote}
% KZP3CDM
The laugh of this participant ridiculed the harmlessness of the corona virus only. The uncertainty about its origin remained.
\par 
\begin{displayquote}... there is a certain group of people who neglect either the mortality or the harmfulness of this coronavirus. Some compare it with the flu. I believe it is more harmful than the flu... (M56) \end{displayquote}
% KWPZOCD
This participant opposed the idea of the usual flu-like harmfulness of the virus and realized that it was more dangerous.\par
We see that when confronted with false news beforehand, the participants' attitude ranged from the unawareness through the uncertainty to the conviction about the information being false.
Due to such a wide spread of awareness of the circulating false information, the need for some kind of support to verify the credibility of COVID-19 news becomes apparent.

%\vspace{-.2cm}
\section{Perceived credibility of Rita}
We were interested in finding out whether the information can be regarded credible when it is provided by the chatbot. Hence, we presented not only proven information about the pandemic, but we also included a section dedicated to correcting of some common myths\footnote{sourced from WHO}. Below comes one of few similar opinions about the credibility of the presented information:

\begin{displayquote}
They [i.e. the news] are very good, comprehensive on every topic (F49) \end{displayquote}
% KZ17IUQ
Generally, the participants were satisfied with the credibility of presented news. However, that credibility was not established by just seeing an innovative technological solution. Rather than that, the credibility was assessed through comparing presented news with user's existing information base or common knowledge. In our context, common knowledge should be understood as a set of news and guidelines widely disseminated by governments and public health authorities, and received through the trusted sources of the respective participant.

\begin{displayquote}At the glance I had, [it is] a rather correct, common knowledge. (F49) \end{displayquote}
% KZ17IUQ
\begin{displayquote} {[it is credible]} when compared with other information (F66) \end{displayquote}
% WMK02

The utterances above show that perceived credibility was assessed through comparison with information that was accepted as credible by the participants beforehand. Interestingly, nobody asked about the origin of any of the presented news. This may be attributable to the fact that no latest news were shown, and the participants had already been familiar with most of the information.

Only one participant had no opinion about Rita's credibility. The reason for this may have been that the person was in search of the latest news only, which were not covered by the application yet. 

The perception of the speaking voicebot is to some extent similar as of a real person. E.g. some users commented on Rita's pitch of voice or demanded to see her as a person:
\begin{displayquote}
I miss her photo. (M51)
\end{displayquote}
% KZ7KU7B

Although we recorded two critical views of Rita, the majority of participants rated Rita positively as a person and assessed some of her personal traits:

\begin{displayquote}
{[Rita's impression is]} very interesting ... and doubtless useful (M70)
\end{displayquote}
% WMK01

\begin{displayquote}
... pleasant and nice, correct, [her] information is comprehensive (F49) \end{displayquote}
% KZ17IUQ

\begin{displayquote}
Rita is well oriented in the matters of the coronavirus. She knows how to recognize it. She can give advice on how to deal with it. (F59)
\end{displayquote}
% KZ97J48

It should be noted that Rita's personal attributes like \textit{professional}, \textit{useful}, \textit{well oriented}, \textit{can give advice} indicate her credibility at a personal level too. Personal credibility of Rita supports the overall perceived credibility of the application.

%\vspace{-.2cm}
\section{Expressing user's emotions}
Respondents appreciated multimodal interface based on emotion wheel. They all selected the emotion on the touch screen, but it was possible to use voice commands. Users felt confident when the Bot understood the input. All of them had experienced errors in other HCI systems based on not correctly recognized user inputs.

\begin{displayquote}
This graph is perfect. It's great, it's very elaborate and everyone can find what they can feel. (F49)
\end{displayquote}

%\vspace{-.2cm}
\section{Factors influencing therapy’s engagement}

A few respondents evaluated sessions as useful and admitted that they helped them to relax, a few regarded it as worthless. Several factors influenced this decision: (1) Due to the limited amount of time, the theoretical framework of the therapy was not introduced to better understanding of the meaning of certain exercises. Increasing the subjects awareness about theoretical background could have positively influenced on the evaluation of the therapy's effectiveness and a sense of each exercise.
Secondly, even if the respondents had had previous experiences of using bots, its purpose was receiving specific information, rather than focusing on activities that required focus and silence. More regular use of Rita could lead to a habituation effect, resulting in concentrating on exercises and effective relationship with the application.
\begin{displayquote}
It makes no sense. (M51)
\end{displayquote}
%KZ7KU7B
\begin{displayquote}
It was relaxing. After the session, you know what and how and you are more focused on getting some messages across. (M47)
\end{displayquote}
%KWG2P5K-OURIVI

A relationship with the therapist built on trust is one of the factors influencing willingness to engage in the therapy. In the study, the role of the therapist was replaced by a multimodal bot, which was perceived by the respondents as a machine. It had a negative impact on building a true and deep therapeutic relationship.
 
 \begin{displayquote}
I think it would be a new thing and it would take a little bit of adjustment, in a sense I rarely talk to the machine. (F50)
\end{displayquote}

The information about effectiveness of the proposed therapeutic techniques could foster trust in the multimodal bot. It would also require a commitment to several therapy sessions.
Respondents declared that they would use a multimodal bot when experiencing anxiety or in the moments of high stress. Interestingly, respondents noted the potential use of a multimodal bots as a possible substitute for relationships for single people
 
\begin{displayquote}
it helps, I could use it. However I have not directly used such methods so far. (M86)
\end{displayquote}
%KZP3CDM

\begin{displayquote}
I suppose that for lonely people or those who don't go to work every day and miss human contact, asking Rita at breakfast in the morning: "How are you feeling? How was your night?" or "How did you get home from work? How was your day?" would be important. (F50) \end{displayquote}
% KWHUE6J
Those observations are the first step to enhance the design of multimodal bot that can be a complement to traditional therapies with psychologists.

\chapter{Conclusion}
We chose to use the qualitative evaluation method as a way to obtain the data about usability as well as the first impressions of the multimodal bot from participants belonging to the target group of users. This shall help us design a quantitative study on a larger group of participants, so that we can prove that a well-designed multimodal bot combining two functions: informative and therapeutic, can be a tool to reducing anxiety caused by pandemic. This goal can be achieved by regaining control over the progress of epidemic and its consequences by following verified guidelines. Moreover, long-lasting use of the implemented therapy module could be an alternative method of anxiety reduction during crisis. 
It was observed that proven news delivered to users by the bot were regarded both helpful and credible. At the same time, when the bot was perceived as a person, that person was regarded credible too.
The quotations from the interviews showed that using multimodal bot to reduce anxiety could be effective, provided that recommendation from the previous sections would be applied, such as: design including explaining the goal of each therapy's steps or information about the effectiveness of a particular therapeutic approach. These guidelines should be applied in both modalities: vocal and visual. \par
The information part plays an important role in establishing trust in the therapeutic part. After learning a lot of information regarded as useful and credible, users learn to trust the bot. Then they will be ready to proceed with the therapeutic function, because they anticipate that whatever the bot proposes next, it can be trusted too. We observe that not only can the bot provide credible information, but doing so is beneficial for building trust in the therapy offered by the bot.
In the future, this solution may be a way of coping with stressful situations caused by unforeseen life and social circumstances such as pandemic outbreak. Especially, the target user group of the multimodal bot may be aging citizens who suffer the most during a crisis.
\par
To further our research we plan to do the quantitative study measuring therapy's efficacy of ACT approach implemented in the multimodal bot. New experiment's approach should include series of therapeutic sessions, so that therapy's healing mechanisms could be established.
\newpage

%\bibliographystyle{abbrv}
%\bibliography{references.bib}
\printbibliography
\end{document}